\documentclass[10pt,a4paper,twocolumn]{article}

\usepackage[right=1.29cm,left=1.29cm,top=1.9cm,bottom=4.29cm]{geometry}
\usepackage{graphicx}
\usepackage{epsfig}
\usepackage{grffile}
\usepackage{lipsum}
\usepackage{subfigure}
\usepackage{color}
\usepackage{times}
\usepackage{amsmath}
\usepackage[authoryear]{natbib}
\usepackage{amsmath,amsfonts,amssymb}
\usepackage{multicol}
\usepackage{multirow}
\usepackage[T1]{fontenc}
\usepackage{url}
\usepackage{array}
\usepackage{titlesec}
\usepackage{indentfirst} 
\usepackage{caption}
\usepackage{authblk}

\titleformat*{\section}{\mdseries\bf\selectfont}
\titleformat*{\subsection}{\mdseries\itshape\selectfont}

\makeatletter
\renewcommand{\@seccntformat}[1]{\csname the#1\endcsname.\quad}
\makeatother

\title{Identification of Credit Risk Based on Cluster Analysis of Account Behaviours}

\author[1, 2]{Maha Bakoben}
\author[1]{Tony Bellotti}
\author[1, 3]{Niall Adams}

\affil[1]{Department of Mathematics, Imperial College London, London SW7 2AZ, United Kingdom}
\affil[2]{Statistics Department, King Abdulaziz University, P.O.Box 80200, Jeddah 21589, Saudi Arabia}
\affil[3]{Heilbronn Institute for Mathematical Research, University of Bristol, Bristol BS8 9AG, United Kingdom} 

\date{}

\begin{document}
\twocolumn[
 \begin{@twocolumnfalse}
 
\maketitle
 \end{@twocolumnfalse}
          ]

\begin{abstract}
Assessment of risk levels for existing credit accounts is important to the implementation of bank policies and offering financial products. This paper uses cluster analysis of behaviour of credit card accounts to help assess credit risk level. Account behaviour is modelled parametrically and we then implement the behavioural cluster analysis using a recently proposed dissimilarity measure of statistical model parameters. The advantage of this new measure is the explicit exploitation of uncertainty associated with parameters estimated from statistical models. Interesting clusters of real credit card behaviours data are obtained, in addition to superior prediction and forecasting of account default based on the clustering outcomes.\\

\end{abstract}


\section{Introduction}
\label{sec1}
Behavioural credit scorecards can be defined as statistical models of customer behaviour, i.e. card usage and repayments, over time \citep{tillandhand}. The aim of these models is to identify which of the existing customers may experience difficulty paying back the loan \citep{thomas02}. Identification of distinct risk levels might support operating decisions with regard to increasing credit limits or offering a financial product \citep{thomas02, tillandhand}. 

In this paper, we present a new methodology for identifying different risk groups based on the available data of customer behaviour. The method aims to assign credit card accounts to clusters such that the behaviours of accounts in the same clusters are similar. This cluster analysis can be used as a tool for building different behavioural scorecards or developing distinct marketing strategies for groups of accounts. 

A typical interest in retail banking is predicting the probability of a customer not being able to make the minimum amount of the agreed monthly repayment for three consecutive months, this event referred to as `default'. A default prediction model based on aggregate summaries of account behaviour is traditionally used in behavioural credit scoring \citep{thomas09}. The aggregate summary can be defined as a statistic which describes the time series in a single value such as the mean or median. This approach might result in loss of valuable information inherent in the dynamic behaviour data. We introduce a new approach to the development of the default prediction and forecasting models. This approach utilises the outcomes of the cluster analysis of the credit behaviours. To distinguish between a {\it prediction} model and a {\it forecasting} model, the former predicts the default status over an observed behaviour period, whereas the forecasting model predicts the default status at a future period after observing the behaviour.  

A fundamental aspect in clustering methods is the specification of a dissimilarity measure that is appropriate for the data. As behaviours can be considered as time series, serial dependence needs to be considered in the definition of the dissimilarity measure.  Two stages for defining the dissimilarity between pairs of time series objects will be considered in this paper. The first is fitting a multivariate time series model to express the dynamic characteristics of the account. This stage reduces the dimension of the data by providing the model parameters as a summary, in addition it makes the dissimilarity comparison feasible between credit accounts with different numbers of transactions.

The second stage computes the dissimilarity between {\it confidence regions} of the model parameters. Since the objects being clustered are parameters of a statistical model, they exhibit statistical uncertainty. Notably, this uncertainty is driven by the amount of data used to estimate the model. This uncertainty-aware dissimilarity measure, recently introduced in \citet{bakoben}, is intended to account for this uncertainty in the estimated model parameters. The consideration of such uncertainty produces more reliable clusters than clusters based only on parameter estimates.  We are not aware of any literature that has addressed the cluster analysis of credit behaviours using the time series clustering approaches that are described in this paper.

A previous study that considers differentiating credit accounts based on their behaviours is the paper by \citet{hsieh}. The author applies a self-organizing map neural network for the purpose of identifying distinct profitable groups based on transaction variables including repayment behaviour. This was based on aggregate values of the account behaviours. In another study by \citet{gao}, credit card accounts were divided into clusters based on an objective cluster analysis (OCA) for application and behavioural variables. The standard Euclidean distance was used for dissimilarity computations. Then a neural network was created for each cluster to predict the `good' and `bad' accounts. Again this study considers aggregate representations of behaviours which may result in loss of valuable information about the dynamic changes in account behaviours over time.

One of the earliest studies concerned with cluster analysis of credit account behaviours is the paper by \citet{edelman} which performs the clustering on delinquency count. In \citet{edelman}, the overall total delinquencies of accounts observed at each month over a two-year period were clustered using the $k$-medoids clustering method with Euclidean distance, where the main purpose of the analysis is to identify clusters of months or a combination of months and products. \citet{adamshand} divide credit card accounts into two clusters on the basis of least squares parameter estimates of a linear regression model. The linear model is fitted to the cumulative numbers of missed repayments over a twelve months period. Again with respect to credit card behaviours, \citet{tillandhand} cluster delinquency counts into groups based on Euclidean distance of the linear slope of a polynomial model for the delinquency count over time. Note that those papers were concerned with clustering a univariate behaviour while the clustering approach presented in this paper is applicable to multiple behaviours.

This paper is organised as follows. Section \ref{sec2} describes the available real data set of credit card account behaviours. Section \ref{sec3} illustrates the two stages of the cluster analysis. Section \ref{sec4} introduces the prediction and forecasting models of default. The empirical results of clustering account behaviours are presented in Section \ref{sec5}. Section \ref{sec6} and \ref{sec7} show the outcomes of the default prediction and forecasting models, respectively. Finally, Section \ref{sec7} summarises the work of this study.

\section{Data set}
\label{sec2}
The credit card data set includes monthly behaviours for 494 accounts at an anonymous bank in the UK for a maximum period of $37$ months. The objective with this data is to assign customers into clusters based on their monthly behaviours and {we aim to discriminate between high and low risk customers}.

For a single customer $s$, we denote the corresponding behavioural credit account by ${\bf Y}_s$ and its length by $T_s$. Each account has the following characteristics: ${\bf y}_{s,\mbox{\scriptsize repay}}$ denotes a vector of the monthly repayment amount made by the customer, ${\bf y}_{s,\mbox{\scriptsize bal}}$ denotes a vector of total balance on the account at the end of each month and ${\bf y}_{s,\mbox{\scriptsize cl}}$ denotes a vector of the monthly credit limit which is static for most customers. The latter two behaviours will be considered indirectly through a new behaviour vector, ${\bf y}_{s,\mbox{\scriptsize ut}}$, that is called \emph{utilisation rate}; the ratio of total balance to credit limit, 
$$
\mbox{utilisation rate}=\frac{\mbox{total balance}}{\mbox{credit limit}},
$$
where the value of utilisation rate should be between $0$ and $1$. However, there are cases when this rate goes below or over the standard range. For example, customers overpay their loans (i.e. ${\bf y}_{s,\mbox{\scriptsize bal}} < 0$) or the total balance exceeds the credit limit (i.e. ${\bf y}_{s,\mbox{\scriptsize bal}} > {\bf y}_{s,\mbox{\scriptsize cl}}$). The mean of utilisation rate in the credit data is $0.6355$. The minimum and maximum values are $-7.0990$ and $3.5600$, respectively.  

Other characteristic in the credit card data set includes delinquency count -- a cumulative number of missed number of payment. This ranges between 0 and 12. In addition, a default status, $x_s(t) \in \{0,1\}$ for $t=1, \ldots T_s$, is defined based on the delinquency count. If a customer misses several consecutive payments (usually three) by time $t$, then the default status $x_s(t)=1$ otherwise $x_s(t)=0$.

Note that only the repayment amount and utilisation rate will be used to build the clusters. 

For the purpose of evaluation, a proportion of $60\%$ training data from the credit accounts is used for the construction of the model and $40\%$ of the accounts are held out for testing. 

\section{Clustering method}
\label{sec3}
This section describes the two stages of defining dissimilarity between credit card behaviours. Section \ref{sub3_1} describes the first stage that is the time series modelling of customer behaviour, Section \ref{sub3_2} describes the conventional approach for defining dissimilarity between model parameters and Section \ref{sub3_3} illustrates the inclusion of parameter uncertainty in the dissimilarity measure for the cluster analysis of credit card behaviours.  

\subsection{Time series modelling}
\label{sub3_1}

First, we reduce the dimension of the observed behaviours to make the dissimilarity comparison feasible between the credit accounts. We follow the time series model-based reduction method in \citet{bakoben15}.

For a single account $s$, the monthly repayment behaviour, ${\bf y}_{s,\mbox{\scriptsize repay}}=[y_{s,\mbox{\scriptsize repay}}\small{(t=1)},\ldots, y_{s,\mbox{\scriptsize repay}}{\small(t=T_s)}]^T$, and the utilisation rate behaviour, ${\bf y}_{s,\mbox{\scriptsize ut}}=[ y_{s,\mbox{\scriptsize ut}}{\small(t=1)}, \ldots, y_{s,\mbox{\scriptsize ut}}{\small(t=T_s)}]^T$, can be described by a bivariate vector autoregression model (VAR) of order one \citep{helmut} as follows:

\begin{equation}
\small{
 {\bf \begin{pmatrix} y_{s,\mbox{\scriptsize repay}}(t)\\ y_{s,\mbox{\scriptsize ut}}(t) \end{pmatrix}
=\begin{pmatrix} \theta_{s,1} \hskip8pt \theta_{s,2} \\ \theta_{s,3} \hskip8pt \theta_{s,4} \end{pmatrix}
\begin{pmatrix} y_{s,\mbox{\scriptsize repay}}(t-1) \\ y_{s,\mbox{\scriptsize ut}}(t-1) \end{pmatrix} + \begin{pmatrix} u_{1}(t) \\ u_{2}(t) \end{pmatrix} },}
\label{ch2:eq14}
\end{equation} 

where $ {{\bf u}=[ u_{1}(t), u_{2}(t)]^T}$ is a vector of weakly stationary white noise process, ${\bf u} \sim N(\bf{0},{\Sigma})$. Each equation in a VAR model is estimated separately by an ordinary likelihood estimator \citep{helmut}. 

By fitting the bivariate VAR model of order one to $N$ behavioural credit accounts, we obtain $N$ vectors of VAR coefficients  $\boldsymbol{\theta}_s=[ \theta_{s,1}, \ldots, \theta_{s,p}]^T$ where in this case $p=4$.

\subsection{Conventional clustering approach}
\label{sub3_2}
As described in \citet{bakoben15}, Euclidean distance can be computed directly between a pair of VAR coefficients vectors. For two credit account behaviours ${\bf Y}_r=[{\bf y}_{r,\mbox{\scriptsize repay}}, {\bf y}_{r,\mbox{\scriptsize ut}}]$ and ${\bf Y}_s=[{\bf y}_{s,\mbox{\scriptsize repay}}, {\bf y}_{s,\mbox{\scriptsize ut}}]$, Euclidean distance between their corresponding VAR coefficients $\boldsymbol{\theta}_r=[\theta_{r,1}, \ldots, \theta_{r,p}]^T$ and $\boldsymbol{\theta}_s=[\theta_{s,1}, \ldots, \theta_{s,p}]^T$ is computed as follows:  \\

\begin{equation}
{ {d_{\mbox{\scriptsize euc}}({\bf Y}_r,{\bf Y}_s)}= \sqrt{\sum_{i=1}^{p} (\theta_{r,i} - \theta_{s,i})^2}.}
\label{eq1}
\end{equation}

\subsection{Uncertainty-aware clustering}
\label{sub3_3}
These estimated VAR parameters are subject to statistical uncertainty. This type of uncertainty can be characterised by the covariance matrix of the estimated parameter vector, denoted by $\Psi$. \cite{bakoben} proposed an approach for the explicit inclusion of uncertainty in the computation of dissimilarity between data points. The idea of the new metric is to measure the overlap between $(1-\alpha)$ confidence regions of VAR coefficients. Each confidence region is represented geometrically by an ellipsoid defined by:

$$ {\small \mathcal{E}_s({\boldsymbol{\theta}_s},{\Psi}_s): \{ ({\bf x}-{\boldsymbol{\theta}_s})^T {(c \hat{\Psi}_s)}^{-1}({\bf x}-{\boldsymbol{\theta}_s}) \le 1 \} },$$ \vskip5pt
\noindent where the scalar $c=\sqrt{{ \scriptsize{p}} F_{\tiny p, T_s-p-1, 1-\alpha}}$, $T_s$ is the length of the corresponding credit account and $\alpha$ is the significance level. 

The ratio of overlap between each pair of ellipsoids $(\mathcal{E}_r,\mathcal{E}_s)$ is given by

\begin{equation}
R_{r,s} \equiv \frac{V_{\mathcal{E}_r \cap \mathcal{E}_s}}{V_{\mathcal{E}_r} + V_{\mathcal{E}_s}- V_{\mathcal{E}_r \cap \mathcal{E}_s}}, \hskip8pt  r \neq s, \hskip5pt V_{\mathcal{E}_r},V_{\mathcal{E}_s} >0,  
\label{eq2}
\end{equation}
\\where the hyper-volumes of ellipsoids $V_{\mathcal{E}_r}$ and $V_{\mathcal{E}_s}$ are computed by the mathematical formula $V_{\mathcal{E}}=\frac{\pi^{p/2} |\Psi|^{1/2}} {\Gamma(p/2 +1)}$ \citep{friendly}. The volume of the overlap region, $V_{\mathcal{E}_r \cap \mathcal{E}_s}$, is estimated by Monte Carlo simulations \citep{robert} as there is no-closed formula for the overlap volume. Then, the dissimilarity between confidence regions of VAR coefficients is defined by

\begin{equation}
d_{\mbox{\scriptsize ell}}({\bf Y}_r,{\bf Y}_s)=1-R_{r,s}, \hskip10pt d_{\mbox{\scriptsize ell}} \in [0,1].
\label{eq3}
\end{equation} 

The next step in the credit behaviours cluster analysis is the implementation of the $k$-medoids partitioning cluster method \citep{kaufman, kaufman09}. Each account is assigned to the cluster with the closest medoid $m$. The $k$-medoids method with the uncertainty-aware dissimilarity attempts to identify clusters that minimise the sum of this distance to the medoids $m_1, \ldots, m_k$. 

For a credit account ${\bf Y}_s$, a vector of cluster allocation ${\bf z}_s=(z_{s,1}, \ldots, z_{s,k})$, is defined where each element in the vector, $z_{s,l}$ for $l=1, \ldots, k$, is given by,  

\begin{equation}
z_{s,l} = \left \{ 
\begin{array}{rcl}
1 & \mbox{if} & \mbox{argmin}_l \hskip3pt d_{\mbox{\scriptsize ell}}({\bf Y}_s, {\bf Y}_{m_l})\\
0 & {} & \mbox{Otherwise}.\\
\end{array} \right. 
\end{equation}

\section{The use of clusters for model predictions and forecasts} 
\label{sec4}
We develop a model to predict default. This model will also be used to evaluate the clustering performance. Here, we introduce a binary response variable, $\tilde{x}_s$, which indicates whether an account has ever been defaulted or not. This binary value for an account $s$ is measured over the available account's period $[t=1, \ldots, t=T_s]$ as follows: 

$$\tilde{x}_s=\mbox{max}[x_{s}(t=1), \ldots, x_{s}(t=T_s)],$$

where $\tilde{x}_s=1$ indicates that the account $s$ has been defaulted at least once. 
 
The default status is predicted based on the cluster assignment that is an explanatory variable in the logistic regression model: 
\begin{equation}
p(\tilde{x}_s=1| {{\bf z}_s})=\frac{e^{\beta_0 + \sum_{j=1}^k \beta_j z_{s,j}}}{1+e^{\beta_0 +\sum_{j=1}^k \beta_j z_{s,j}}}.
\label{eq4}
\end{equation}

\begin{figure}[h!]
\centering
\includegraphics[width=3.5in,trim={140 622 150 100},clip,scale=1.8]{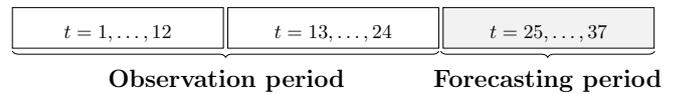}
\caption{{\small Illustration for observation and forecasting period in default forecasting model.}}
\label{fig1}
\end{figure}

Equation \ref{eq4} is also used for forecasting in which the cluster analysis is performed on the first 2/3 of each account profile and the forecast default is measured over the last 1/3 period. This is due to variable lengths of the available credit account profiles, hence choosing specific lengths for the observation and forecast periods is not reasonable as the profile length of an account might be less than the specified observation period. Figure \ref{fig1} illustrates the observation and forecast period in the default forecasting model. Note that some time-window after credit card origination is required to allow for measurement in observation period (e.g. $t=13$ to $t=24$ in Figure \ref{fig1}). 

In order to evaluate the new default model, we compare its performance to the conventional aggregate model. This models the default status with aggregate representations of the time series defined here by a vector ${\bf g}_s=(\bar{g}_{\mbox{\scriptsize repay}}, \bar{g}_{\mbox{\scriptsize ut}})^T$ which consists of the mean values of the univariate time series ${\bf y}_{s,\mbox{\scriptsize repay}}$ and ${\bf y}_{s,\mbox{\scriptsize ut}}$. The aggregate repayment behaviour is given by 

\begin{equation}
\bar{g}_{\mbox{\scriptsize repay}}= \frac{\sum_{t=1}^{T_s} y_{s,\mbox{\scriptsize repay}}(t)}{T_s}. 
\label{eq5}
\end{equation}

Similarly, the aggregate utilisation rate behaviour is computed. Then, the aggregate default model is defined by

\begin{equation}
{ p(\tilde{x}_s = 1| {\bf g}_s) =\frac{e^{\beta_0 + {\boldsymbol \beta}.{\bf g}_s }}{1+ e^{\beta_0 + {\boldsymbol \beta}.{\bf g}_s}}},
\label{eq6}
\end{equation}

where ${\boldsymbol \beta}$ is a $2$-dimensional parameter vector for the aggregate representations ${\bf g}_s$. 

The prediction and forecasting performance of default models are evaluated by the following common assessment criteria: the H-measure \citep{davidhand}, Kolmogorov-Smirnov statistic \citep{duda}, Gini-index \citep{tibshirani} and area under the receiver-operating characteristic curve (AUC) \citep{fawcett}.

\section{Results: Clusters of credit card behaviours}
\label{sec5}
We apply the uncertainty-aware clustering method described in Section \ref{sec3} with the number of clusters $k=3$. This number was selected considering a reasonable objective clustering of the credit account behaviours and finding the optimal number of clusters is beyond the scope of this paper.

The proportions of credit accounts in the three clusters are presented in Table \ref{tab1}. Table \ref{tab1} also shows the cluster outcomes based on clustering using the standard Euclidean distance. In comparison to the ellipsoid based clusters, Euclidean distance tends to create one cluster that includes a large proportion of accounts whereas the other clusters include a small proportion of the credit account sample. For example, cluster $\mathcal{C}_1$ comprises $62\%$ of the total accounts. This demonstrates the importance of incorporating uncertainty in clustering.

\begin{table}[h!]
\centering
\caption{{\small The size of clusters $\mathcal{C}_1, \mathcal{C}_2$ and $\mathcal{C}_3$ obtained from the $k$-medoids clustering using the ellipsoid dissimilarity $d_{\mbox{\scriptsize ell}}$ and Euclidean distance $d_{\scriptsize \mbox{euc}}$ for a sample of $494$ credit accounts.}}
\begin{tabular}{lccc}
  \hline
$k$-medoids cluster $\#$ & $\mathcal{C}_1$ & $\mathcal{C}_2$ & $\mathcal{C}_3$ \\ 
  \hline
$d_{\mbox{\scriptsize ell}}$ & $244(50\%)$ & $115(23\%)$ & $135(27\%)$ \\ 
$d_{\scriptsize \mbox{euc}}$ & $307(62\%)$ & $144(29\%)$ &  $43(9\%)$ \\ 
   \hline
\end{tabular}
\label{tab1}
\end{table}

\begin{figure*}[h!t]
\centering
$\begin{array}{cc}
\includegraphics[width=2.7in]{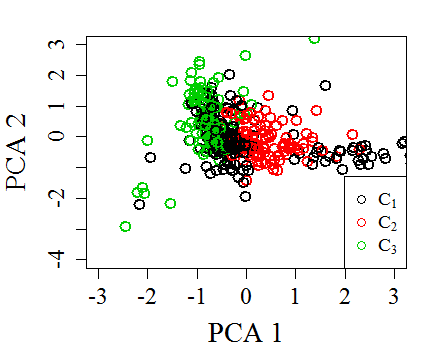}
\includegraphics[width=2.7in]{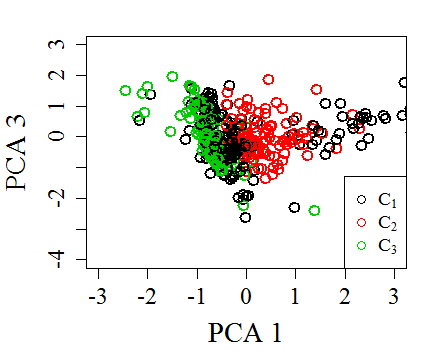}\\
\includegraphics[width=2.7in]{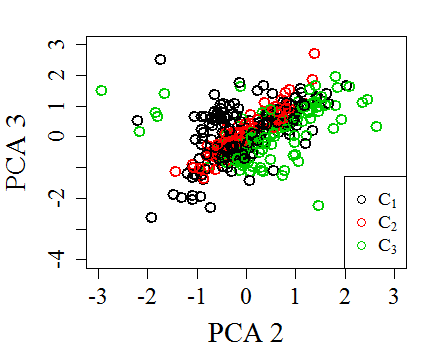}
\end{array}$
\caption{{\small Visualising the outcomes of the $k$-medoids clustering based on the ellipsoid dissimilarity $d_{\mbox{\scriptsize ell}}$ for VAR parameter estimates, using principal components.}}
\label{fig2}
\end{figure*}

The outcome of the $k$-medoids clustering is displayed in Figure \ref{fig2} using principal component analysis (PCA) \citep{pca}. The PCA was applied to the VAR parameter estimates, ${\boldsymbol{\theta}}$, to visualise the structure of the resulting clusters by reducing the dimensionality of the VAR parameters. The PCA plots show major trends which are determined by the first three components. It appears that the three clusters are not clearly separated and it is difficult to observe clear patterns in the original parameter space.  

To further investigate the behaviours of the accounts in the three clusters, we study each behaviour in the data space.

\subsection*{Repayment amount, credit limit and total balance behaviours}
\begin{figure*}[h!t]
\centering
$\begin{array}{cc}
\includegraphics[width=2.7in,page=4]{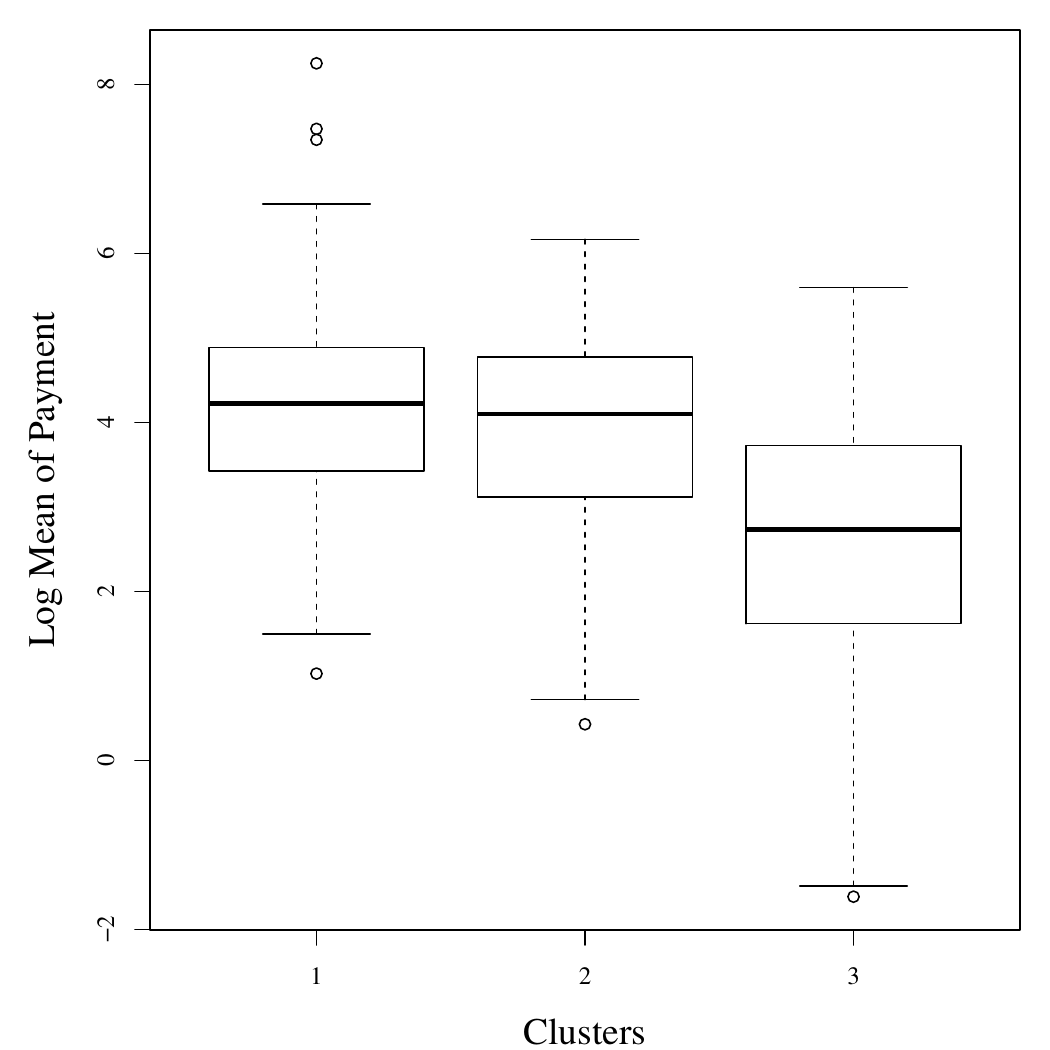}
\includegraphics[width=2.7in,page=5]{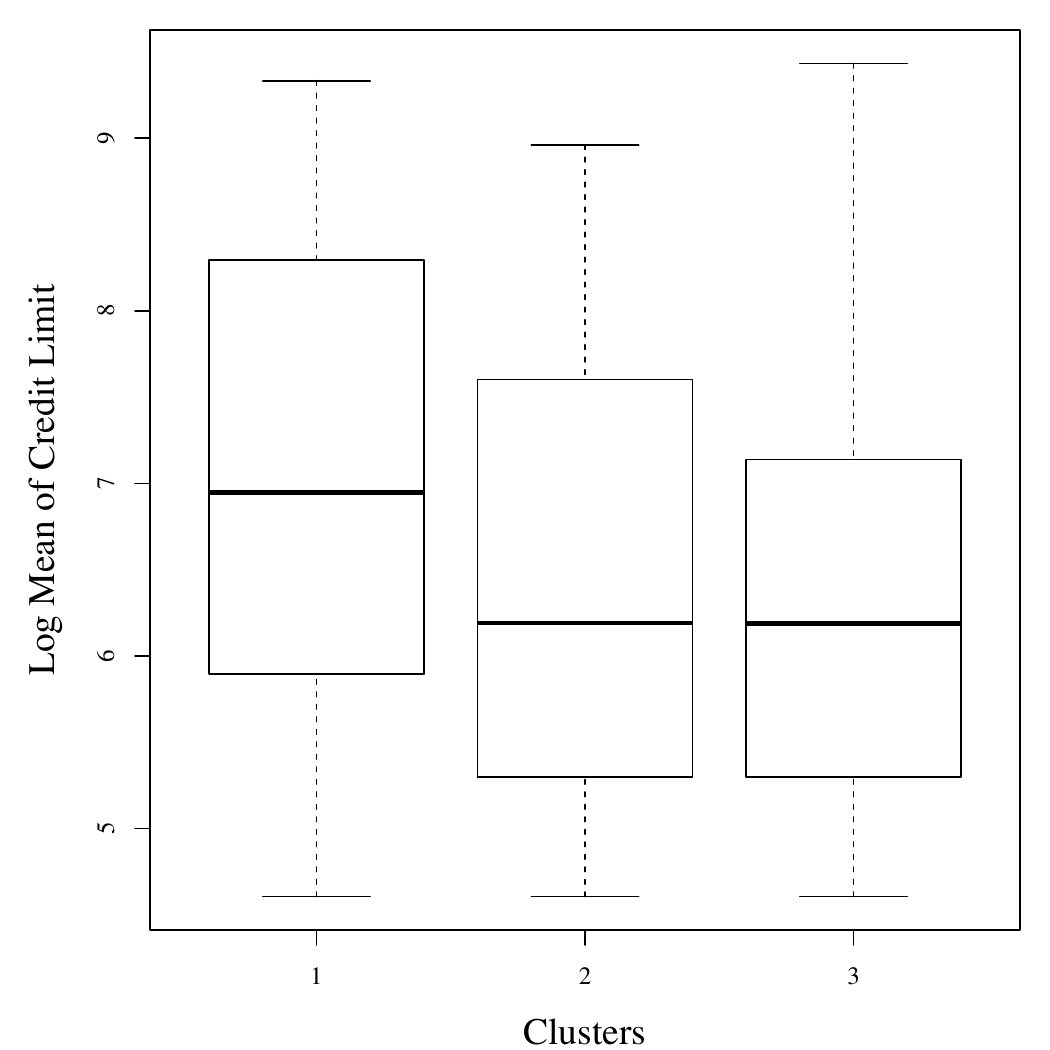}\\
\includegraphics[width=2.7in,page=6]{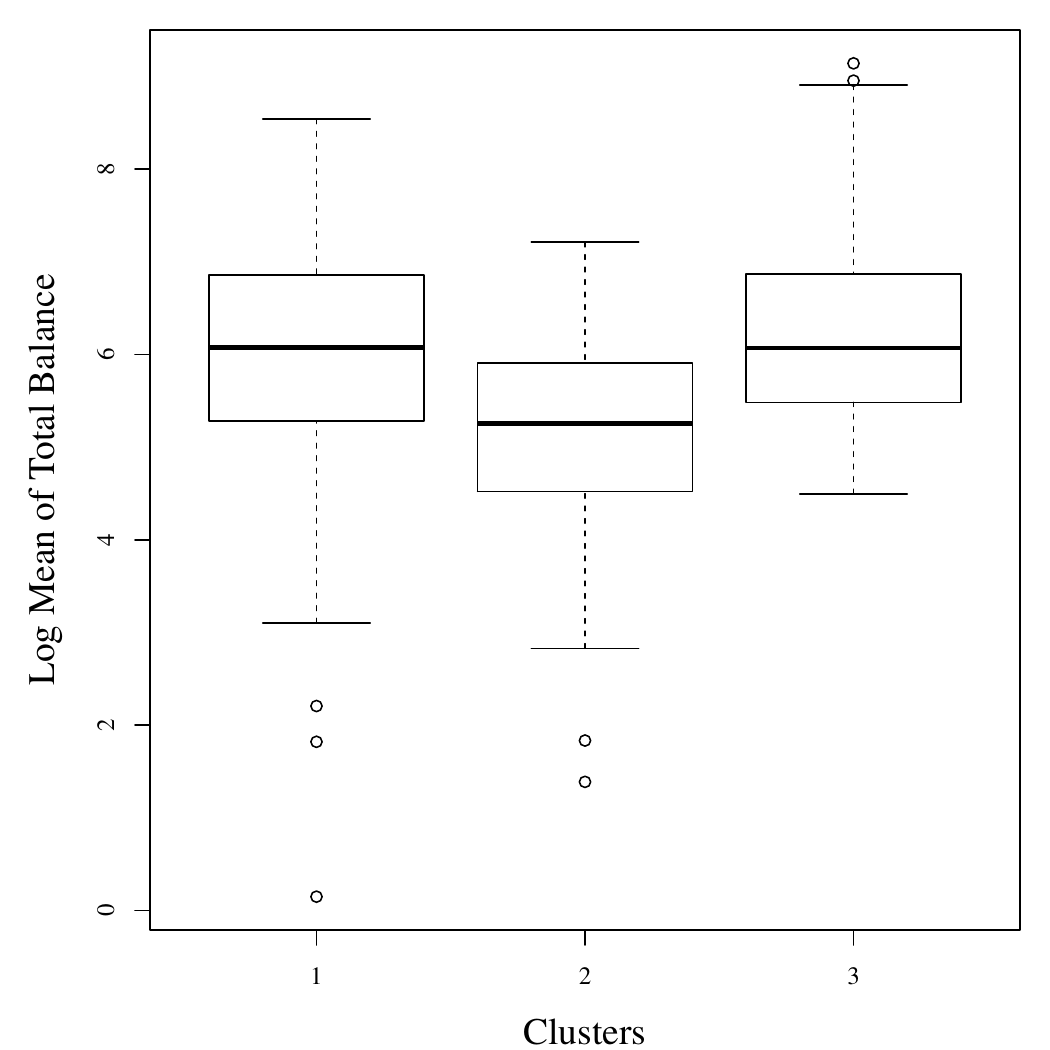}
\end{array}$
\caption{{\small Behaviours of accounts at clusters $\mathcal{C}_1, \mathcal{C}_2$ and $\mathcal{C}_3$. These behaviours are measured on different scales. The clustering results are obtained from the $k$-medoids method with the ellipsoid dissimilarity $d_{\mbox{\scriptsize ell}}$ applied to VAR parameter estimates for credit account behaviours data set.}}
\label{fig3}
\end{figure*}

The boxplots in Figure \ref{fig3} show representations of the account behaviours in data space on the basis of the outcomes of uncertainty-aware clustering which was performed on the parameter space. These plots represent the logarithm of the behaviour sample means for the account profiles separately for clusters $\mathcal{C}_1$, $\mathcal{C}_2$ and $\mathcal{C}_3$.

The credit accounts belonging to cluster $\mathcal{C}_3$ seem to make low payments compared to accounts in the other two clusters. This amount is slightly larger in cluster $\mathcal{C}_1$ than $\mathcal{C}_2$. As shown in the second plot, the highest credit limit seems to be for the accounts assigned to cluster $\mathcal{C}_1$, whereas almost equal median values of the credit limits are observed for the accounts in clusters $\mathcal{C}_2$ and $\mathcal{C}_3$. Although the median of credit limits for the accounts in cluster $\mathcal{C}_3$ is lower than the median for accounts in $\mathcal{C}_1$, both groups seem to have equal amount of outstanding debt (Total balance). That might be because they spend equal amount of money or members of cluster $\mathcal{C}_3$ are not paying their debt or paying only small amount of the debt. This information can be explored by comparing the boxplots of the total balance and credit limits between the three clusters. Additionally, a few extremely low outstanding amounts in $\mathcal{C}_1$ and high outstanding amounts in $\mathcal{C}_3$ are clearly observed.  

\subsection*{Delinquency Counts}
\begin{figure*}[h!t]
\centering
$\begin{array}{cc}
\includegraphics[width=2.7in,page=1]{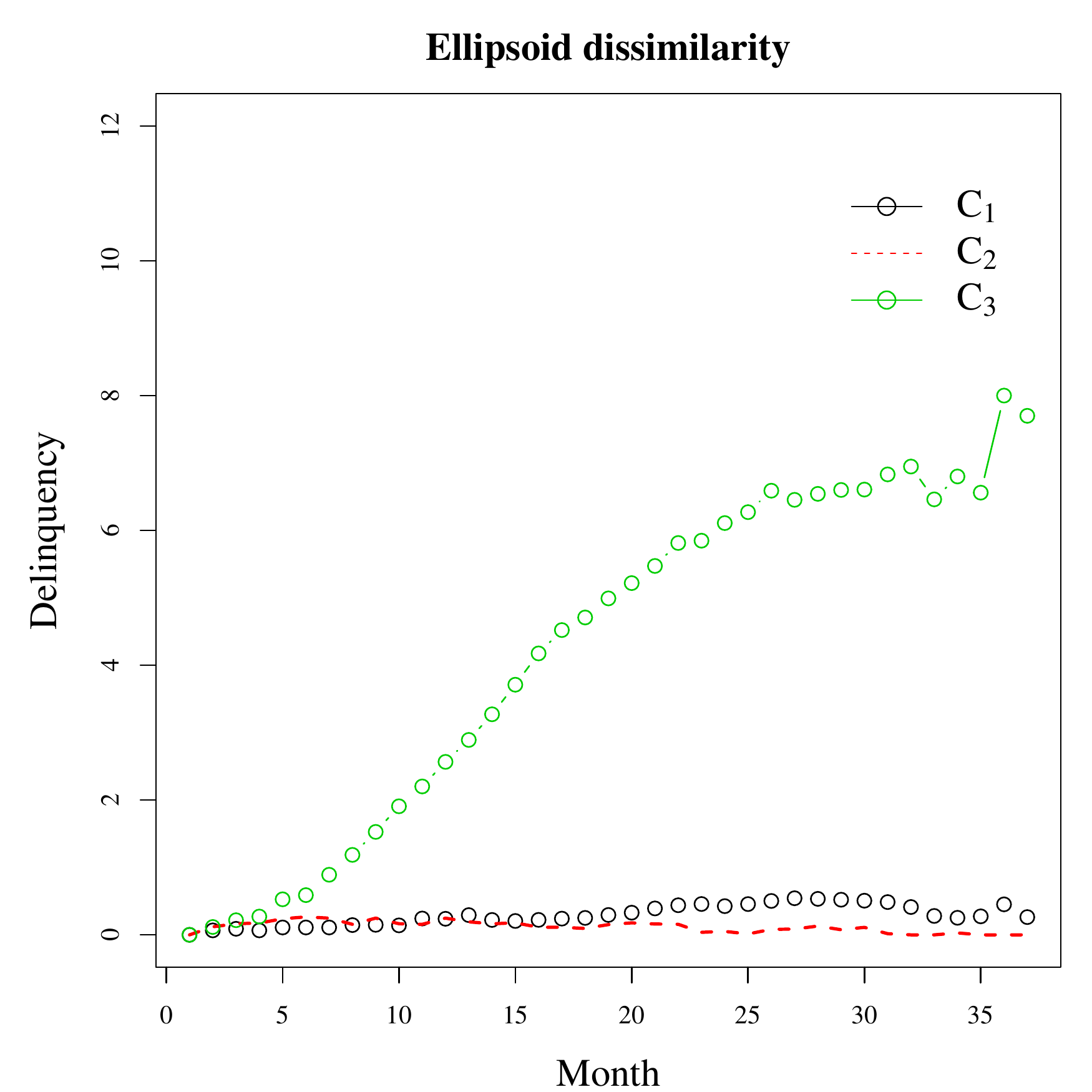}
\includegraphics[width=2.7in,page=2]{summary_delinquency}
\end{array}$
\caption{{\small The means of delinquency counts in clusters $\mathcal{C}_1, \mathcal{C}_2$ and $\mathcal{C}_3$ over the available $37$ months of account records. Clusters are obtained using the $k$-medoids method with the ellipsoid dissimilarity $d_{\mbox{\scriptsize ell}}$ (left plot) and Euclidean distance $d_{\scriptsize \mbox{euc}}$ (right plot). The cluster analysis was applied to the VAR parameters of $494$ accounts.}}
\label{fig4}
\end{figure*}

In this section, we explore the delinquency behaviour in the obtained clusters. Recall this variable was not included in the clustering process but the cluster analysis of VAR parameters revealed interesting aspects of the delinquency behaviour as shown in Figure \ref{fig4}. This figure represents the sample means of delinquency for the credit accounts within each cluster at each month from $t=1$ to $t=37$. 

From the left plot in Figure \ref{fig4}, the credit accounts in the three clusters might be described as those accounts who never default as the delinquency count is always less than 2. These accounts were assigned to cluster $\mathcal{C}_1$. Also accounts in $\mathcal{C}_2$ never defaulted and the mean of the delinquency count for this group is less than the mean of delinquency for accounts in cluster $\mathcal{C}_1$ particularly when $t>20$. In contrast, the last cluster $\mathcal{C}_3$ includes those whose delinquency count is gradually increasing over their profile period and consequently default. Thus, cluster $\mathcal{C}_3$ can be considered as the highest risk group compared to the other two clusters.

Looking at the delinquency plot obtained from the clustering approach that utilises Euclidean distance (right plot in Figure \ref{fig4}), the general structure seems to be similar to the structure obtained using the ellipsoid dissimilarity measure. However, the Euclidean based clustering approach seems to assign some of the high risk accounts to the other clusters. This is clearly apparent by comparing the overall means of the delinquency over time in the high risk cluster in Euclidean clustering and these measured based on the clustering outcomes of the ellipsoid dissimilarity measure. 

\section{Default prediction model}
\label{sec6}
In this section, we present the result of fitting a logistic regression model for default based on the outcomes of clustering VAR model parameters taking into account the associated error estimates. The default status is computed over the available profile period of each account. This prediction model was previously defined in Equation \ref{eq4}, in which the cluster assignment is a predictor variable for the binary default status. In this section, the prediction model is fitted to a training sample of size $296$ credit accounts.

This model is compared to the default prediction model in which cluster analysis was performed without the consideration of the VAR parameter errors. The performance is measured for a test sample of size $198$ credit accounts.

Table \ref{tab2} reports the coefficient estimates of the default prediction models for the $k$-medoids clustering outcomes. It is interesting to find that the cluster assignment is statistically significant for predicting the default status in most cases. Different signs for the influence of the cluster assignment on the default status are only observed in the clusters using the ellipsoid dissimilarity measure $d_{\mbox{\scriptsize ell}}$. The coefficient associated with cluster $\mathcal{C}_2$ suggests a negative effect of the cluster assignment on the default status relative to cluster $\mathcal{C}_1$, whereas cluster $\mathcal{C}_3$ suggests a positive effect relative to $\mathcal{C}_1$. This indicates the proportion of the default class in one of the clusters is higher than in other clusters. A noticeable negative coefficient with high error is observed for cluster $\mathcal{C}_3$ when clustering using Euclidean distance. This is due to the small number of objects in this cluster of which none are from the default class (see Table \ref{tab3}).

Table \ref{tab3} displays the frequencies and proportions of default/non-default across clusters in the training data. The results of ellipsoid dissimilarity measure show the proportion of default in cluster $\mathcal{C}_3$ is relatively higher than the other two clusters, cluster $\mathcal{C}_1$ comes next and the lowest proportion is accounted for cluster $\mathcal{C}_2$. 

\begin{table*}[h!t]
\centering
\caption{{\small Coefficients of the logistic regression models for predicting account defaults based on cluster assignments. Cluster analysis is performed using the $k$-medoids method with the proposed ellipsoid dissimilarity measure $d_{\mbox{\scriptsize ell}}$ and Euclidean distance $d_{\scriptsize \mbox{euc}}$. The regression models are built on a training sample. Note that $\mathcal{C}_1$ is the baseline category.}}
\small
\begin{tabular}{lrrrr}
  \hline
$k$-medoids method & Estimate & Std. Error & z value & $p(|z|)$ \\ 
  \hline
{\bf (a) Ellipsoid dissimilarity $d_{\mbox{\scriptsize ell}}$}& & & &\\
Intercept & $-1.6964$ & $0.2319$ & $-7.3147$ & $2.58e^{-13}$ \\ 
$\mathcal{C}_1$ & - & - & - & - \\
$\mathcal{C}_2$ & $-0.5473$ & $0.4602$ & $-1.1893$ & $0.2343$ \\ 
$\mathcal{C}_3$ & $2.3345$ & $0.3292$ & $7.0916$ & $1.33e^{-12}$ \\ 
\hline
{\bf (b) Euclidean distance $d_{\scriptsize \mbox{euc}}$}& & & &\\
Intercept & $-0.4647$ & $0.1514$ & $-3.0686$ & $0.0022$ \\ 
$\mathcal{C}_1$ & - & - & - & - \\
$\mathcal{C}_2$ & $-1.5068$ & $0.3557$ & $-4.2366$ & $2.27e^{-5}$ \\ 
$\mathcal{C}_3$ & $-17.1014$ & $843.4605$ & $-0.0203$ & $0.9838$ \\ 
\hline

\end{tabular}
\label{tab2}
\end{table*}

\begin{table}[ht]
\centering
\caption{{\small Frequencies of default/non-default status in a training sample. Clusters are obtained using the $k$-medoids clustering method with the ellipsoid dissimilarity measure $d_{\mbox{\scriptsize ell}}$ and Euclidean distance $d_{\mbox{\scriptsize euc}}$.}}
\begin{tabular}{lccc}
  \hline
$k$-medoids clusters & $\mathcal{C}_1$ & $\mathcal{C}_2$ & $\mathcal{C}_3$\\ 
  \hline
$d_{\mbox{\scriptsize ell}}$& & & \\  
non-default($\tilde{x}=0$) & $127(43\%)$  &  $62(21\%)$ &  $29(10\%)$ \\ 
default(${\tilde x}=1$) &  $19(6\%)$ &   $8(3\%)$ &  $51(17\%)$ \\ 
\hline
$d_{\scriptsize \mbox{euc}}$& & & \\ 
non-default(${\tilde x}=0$) & $117(40\%)$ &  $76(26\%)$ &  $25(9\%)$ \\ 
default(${\tilde x}=1$) &  $67(23\%)$ &  $11(4\%)$ &   $0(0\%)$ \\ 
   \hline
\end{tabular}
\label{tab3}
\end{table}

A comparison to the aggregate model (Equation \ref{eq6}) for the prediction performance of account defaults is reported in Table \ref{tab4}. It is interesting to find that including the uncertainty of the statistical model parameters in the cluster analysis improved the prediction performance of the default status with AUC value of $0.7637$ (s.e. $0.0397$). The prediction model using the cluster assignment based on the ellipsoid dissimilarity measure also performs well in comparison to the aggregate model, AUC $0.5310$ (s.e. $0.0450$). 

Additional models are created in an attempt to improve the default prediction performance. Both the cluster assignment and the aggregate behaviours are included as predictors in the logistic regression model. Note that, adding the cluster assignment variable to the aggregate model shows a remarkable improvement in the default prediction performance and the uncertainty clustering made an even bigger improvement to the aggregate model. 

\begin{table}[h!]
\centering
\caption{{\small Performance assessments for default prediction models based on the cluster assignment obtained from the $k$-medoids clustering method using the Ellipsoid dissimilarity $d_{\mbox{\scriptsize ell}}$ and the standard Euclidean distance $d_{\scriptsize \mbox{euc}}$. These models are compared with the default prediction model based on aggregate means of the behaviours. The assessment is performed on a hold-out test sample.}}
\small
\begin{tabular}{lcccc}
  \hline
 & H-measure & KS & Gini & AUC \\ 
\hline
$d_{\mbox{\scriptsize ell}}$  & {0.3748} & {0.5443} & {0.5273} & {0.7637} \\ 
$d_{\scriptsize \mbox{euc}}$ & 0.1416& 0.3405 &0.3563& 0.6781 \\ 
\hline
Aggregate model &0.0573& 0.1569& 0.0620& 0.5310 \\ 
\hline
Aggregate model+  $d_{\mbox{\scriptsize ell}}$  & 0.3962& 0.5543& 0.5110 &0.7555 \\ 
Aggregate model+  $d_{\scriptsize \mbox{euc}}$  & 0.1818 &0.3679 &0.3527 &0.6764 \\  
\hline
\end{tabular}
\label{tab4}
\end{table}

\section{Default forecasting model}
\label{sec7}
This section focuses on forecasting the default status of the credit accounts over unseen future periods of their profiles. As with the prediction models presented in the previous section, the cluster assignment is used as an explanatory variable in the forecasting models. Recall the clusters are obtained from the observation period, whereas the default is computed over the forecasting period. Table \ref{tab5} presents the frequencies of default/non-default classes in the training data.

\begin{table}[h!]
\centering
\caption{{\small Frequencies of default/non-default status in a training sample for the default forecasting model. Clusters are obtained using the $k$-medoids clustering method with the ellipsoid dissimilarity measure $d_{\mbox{\scriptsize ell}}$ and Euclidean distance $d_{\scriptsize \mbox{euc}}$.}}
\begin{tabular}{lccc}
  \hline
Clusters & $\mathcal{C}_1$ & $\mathcal{C}_2$ & $\mathcal{C}_3$\\ 
  \hline
$d_{\mbox{\scriptsize ell}}$& & & \\  
non-default(${\tilde x}=0$) & $165 (30\%)$ & $127 (23\%)$ & $114 (20\%)$ \\ 
default(${\tilde x}=1$) &   $6 (1\%)$ &  $30 (5\%)$ & $119 (21\%)$ \\ 
\hline
$d_{\scriptsize \mbox{euc}}$& & & \\ 
non-default(${\tilde x}=0$) & $284 (51\%)$ & $110 (19\%)$ &  $12 (2\%)$ \\ 
default(${\tilde x}=1$) & $152 (27\%)$ &   $3 (1\%)$ &   $0 (0 \%)$ \\ 
\hline
\end{tabular}
\label{tab5}
\end{table}

Table \ref{tab6} reports the coefficients of the logistic regression models for forecasting the default status based on the cluster assignments. Interestingly, the coefficient estimates are only significant in the model based on the uncertainty-aware dissimilarity measure. Again, as observed in the prediction model in the previous section, some of the forecasting model coefficients have high standard errors as a result of small samples of the default class. 

\begin{table*}[ht]
\centering
\caption{{\small Coefficients of the logistic regression models for default forecasting based on cluster assignments. Cluster analysis is performed using the $k$-medoids method with the ellipsoid dissimilarity measure $d_{\mbox{\scriptsize ell}}$ and Euclidean distance $d_{\scriptsize \mbox{euc}}$. The regression models are built on the training sample. The baseline cluster is $\mathcal{C}_1$.}}
\small
\begin{tabular}{lrrrr}
  \hline
$k$-medoids method & Estimate & Std. Error & z value & $p(|z|)$ \\ 
  \hline
{\bf (a) Ellipsoid dissimilarity $d_{\mbox{\scriptsize ell}}$}& & & &\\
Intercept & $-3.1209$ & $0.4171$ & $-7.4827$ & $7.28e^{-14}$ \\ 
$\mathcal{C}_1$ & - & - & - & - \\
$\mathcal{C}_2$ & $1.8766$ & $0.4463$ & $4.2044$ & $2.62e^{-5}$ \\ 
$\mathcal{C}_3$ & $2.8363$ & $0.4420$ & $6.4175$ & $1.39e^{-10}$ \\ 
\hline
{\bf (b) Euclidean distance $d_{\scriptsize \mbox{euc}}$}& & & &\\
Intercept & $-0.7726$ & $0.1026$ & $-7.5342$ & $4.91e^{-14}$ \\ 
$\mathcal{C}_1$ & - & - & - & - \\
$\mathcal{C}_2$ & $-17.7934$ & $639.5973$ & $-0.0278$ & $0.9778$ \\ 
$\mathcal{C}_3$ & $-17.7934$ & $1581.9722$ & $-0.0112$ & $0.9910$ \\ 
\hline
\end{tabular}
\label{tab6}
\end{table*}

Table \ref{tab7} compares the forecasting performance between the proposed forecasting models, where the cluster assignment is the explanatory variable. Similar to the prediction model, the proposed forecasting models are compared to the forecast model on aggregate summaries. The most favourable model is that based on clustering VAR parameters with the associated uncertainty. The performance values might be reasonable for this particular type of application. The AUC for the best model is $0.7251$ (s.e. $6 \times 10^{-4}$), whereas the AUC for the model based on Euclidean distance is $0.6123$ (s.e. $4 \times 10^{-4}$). The forecast model based on aggregate summaries shows the lowest performance, AUC $0.5355$ (s.e. ${7 \times 10^{-4}}$). 
 
\begin{table}[hb]
\centering
\caption{{\small Performance assessments for default forecasting model based on the cluster assignments obtained from the $k$-medoids clustering method using the Ellipsoid dissimilarity measure $d_{\mbox{\scriptsize ell}}$ and the standard Euclidean distance $d_{\scriptsize \mbox{euc}}$. These models are compared with the default forecasting model based on aggregate means of the behaviours. The assessment is performed on the test sample.}}
\small
\begin{tabular}{lcccc}
  \hline
 & H-measure & KS & Gini & AUC \\ 
\hline
$d_{\mbox{\scriptsize ell}}$  & {0.1825}  & {0.3382} & {0.4502} & {0.7251}\\ 
$d_{\scriptsize \mbox{euc}}$ & 0.0810  & 0.2237 & 0.2246 & 0.6123  \\ 
\hline
Aggregate model  & 0.0276  & 0.1219 & 0.0709  & 0.5355 \\ 
\hline
Aggregate model +  $d_{\mbox{\scriptsize ell}}$  & 0.1744  & 0.3230 & 0.4094  & 0.7047  \\ 
Aggregate model +  $d_{\scriptsize \mbox{euc}}$  & 0.1130  & 0.2799 & 0.2172 & 0.6086 \\  
\hline
\end{tabular}
\label{tab7}
\end{table}

\section{Conclusion}
\label{sec8}
This paper introduced a new behavioural clustering approach that can support the construction of behavioural credit scorecards. In the clustering process, the credit accounts were represented by statistical parameter estimates of their behaviours to represent their associated serial dependence. This results in a significant dimension reduction of data. In addition, the uncertainty of the parameter estimates was considered using an uncertainty-aware dissimilarity measure. 

Taking into account the uncertainty of the model parameters has revealed interesting behavioural clusters. Although the delinquency behaviour of the accounts was not included in the clustering process, the cluster analysis was able to differentiate between the high risk group and low risk group.

We also developed a new default model that includes cluster assignments that can be used for prediction and forecasting purposes. This models default status with cluster assignments. Both the prediction and forecasting models based on the ellipsoid dissimilarity clusters have shown good performance in comparison to the models based on the outcomes of the cluster analysis which ignores the uncertainty of the parameters, and also better performance than models based on aggregate behaviours. 

This research could be extended by performing the uncertainty cluster analysis on time-windows over the account profiles and study the changes in risk levels over profile history. This is an interesting extension of the study but it requires longer behaviour profiles.

\section*{Acknowledgments}
This work was supported by King Abdulaziz University scholarship fund.

\bibliographystyle{apalike}
  \bibliography{bakoben}

\end{document}